\documentclass{article}
\usepackage{spconf,amsmath,graphicx}

\usepackage{subfigure}

\usepackage{float}
\restylefloat{table}
\usepackage{multirow,booktabs}
\usepackage{multicol}

\usepackage{enumitem}
\usepackage{url}
\usepackage{hyperref} 
\usepackage{pdfpages}

\title{EFFECTIVENESS OF 3VQM IN CAPTURING DEPTH INCONSISTENCIES}
\name{  Dogancan Temel and Ghassan AlRegib }
\address{School of Electrical and Computer Engineering, Georgia Institute of Technology\\
            Atlanta, GA, 30332-0250 USA\\
           \{cantemel, alregib\} @ gatech.edu}
\begin{document}

\onecolumn % make sure you keep this coverpage as one column. In this template, we force the coverpage to be one column with this command and then switch to double column for the remaining of the paper with the \doublecolumn command. 

\begin{description}[labelindent=1cm,leftmargin=3cm,style=multiline]

\item[\textbf{Citation}]{D. Temel and G. AlRegib, "Effectiveness of 3VQM in capturing depth inconsistencies," IVMSP 2013, Seoul, 2013, pp. 1-4.} \\

\item[\textbf{DOI}]{\url{https://doi.org/10.1109/IVMSPW.2013.6611918}} \\

\item[\textbf{Review}]{Date added to IEEE Xplore: 26 September 2013} \\

\item[\textbf{Slides}]{\url{https://ghassanalregib.com/publications/}} \\

\item[\textbf{Bib}] {
@INPROCEEDINGS\{Temel2013\_IVMSP,\\ 
author=\{D. Temel and G. AlRegib\},\\ 
booktitle=\{IVMSP 2013\},\\ 
title=\{Effectiveness of 3VQM in capturing depth inconsistencies\},\\ 
year=\{2013\},\\
pages=\{1-4\},\\ 
doi=\{10.1109/IVMSPW.2013.6611918\},\\ 
month=\{June\},\}\\
} \\

\item[\textbf{Copyright}]{\textcopyright 2013 IEEE. Personal use of this material is permitted. Permission from IEEE must be obtained for all other uses, in any current or future media, including reprinting/republishing this material for advertising or promotional purposes,
creating new collective works, for resale or redistribution to servers or lists, or reuse of any copyrighted component
of this work in other works. } \\

\item[\textbf{Contact}]{\href{mailto:alregib@gatech.edu}{alregib@gatech.edu}~~~~~~~\url{https://ghassanalregib.com/}\\ \href{mailto:dcantemel@gmail.com}{dcantemel@gmail.com}~~~~~~~\url{http://cantemel.com/}}
\end{description} 

\thispagestyle{empty}
\newpage
\clearpage

\twocolumn

%\ninept
%
\maketitle
\begin{abstract}
The 3D video quality metric (3VQM) was proposed to evaluate the temporal and spatial variation of the depth errors for the depth values that would lead to inconsistencies between left and right views, fast changing disparities, and geometric distortions. Previously, we evaluated 3VQM against subjective scores. In this paper, we show the effectiveness of 3VQM in capturing errors and inconsistencies that exist in the rendered depth-based 3D videos. We further investigate how 3VQM could measure excessive disparities, fast changing disparities, geometric distortions, temporal flickering and/or spatial noise in the form of depth cues inconsistency. Results show that 3VQM best captures the depth inconsistencies based on errors in the reference views. However, the metric is not sensitive to depth map mild errors such as those resulting from blur. We also performed a subjective quality test and showed that 3VQM performs better than PSNR, weighted PSNR and SSIM in terms of accuracy, coherency and consistency.
\end{abstract}

\section{Introduction}
\label{sec:intro}

Compared to 2D video, Three-Dimensional Television (3DTV) and Free Viewpoint Video (FVV) provide a realistic experience to the user by simulating the depth perception. Furthermore, FVV provides an interactive experience by enabling the user to navigate through the scene. In order to support depth perception, the overall system of 3D from content generation to display differs from existing 2D standards. In stereoscopic 3D (S3D) systems, we need both right and left views corresponding to a specific viewpoint. However, it is not feasible and not always possible to locate stereo camera systems at every single point that we would like to capture the scene. Moreover, as the number of viewpoints increase, 3D system will require higher computational capabilities and more complex coding and streaming techniques. To overcome feasibility issues and practical limits, researchers and developers have worked on various techniques such as depth image-based rendering (DIBR)~\cite{Fehn04}.

Using DIBR, we need a single reference view and the corresponding depth map to synthesize a virtual view at new viewpoints. From content generation to the display side, each of the steps in the DIBR-based 3DTV processing chain affect the perceived quality. The technical report in~\cite{Boev08} describes and categorizes stereoscopic artifacts that can occur inside a 3DTV processing chain. Performance of 3DTV and FVV systems are tested based on these artifacts in the literature. Compression and transmission contribute to the artifacts as outlined in~\cite{Joveluro10} and~\cite{Yunqiang10}. It is established in the literature that the quality assessment of 3D videos inherently differs from quality assessment of 2D. As a result, researchers worked on 3D-specific concepts such as \emph{naturalness} and \emph{viewing experience} under varying blur and depth levels \cite{Kaptein08}. A broader discussion about challenges and advances in multimedia quality assessment can be found in \cite{MQA2011}.

In this paper, we specifically discuss the effectiveness of 3VQM in capturing certain types of distortions. The distortions we consider in this paper are limited to blur, compression artifacts, transmission losses and depth map estimation errors. We start by summarizing 3VQM in Section~\ref{sec:3vqm}. Then, we describe the distortion types and analyze the performance of 3VQM in Section~\ref{sec:performance}. We focus on validation in Section~\ref{ssec:validation} and conclude the paper in Section~\ref{sec:conc}.

\section{A 3D Video Quality Measure (3VQM)}
\label{sec:3vqm}

\emph{3VQM} is a 3D Video Quality Measure that objectifies the visual discomfort in the stereoscopic videos. We obtain \emph{3VQM} by combining three distortion measures defined as spatial outliers (\emph{SO}), temporal outliers (\emph{TO}) and temporal inconsistencies (\emph{TI}). We will briefly describe these distortion metrics but readers are encouraged to look at~\cite{Solh2011} for a detailed description.

Depth maps may not be accurate because of errors in depth estimation, rounding, compression and transmission. Therefore, we need to define an ideal depth map that would generate a visual distortion-free 3D video using DIBR. This definition implies that the video is free from DIBR-induced excessive disparities, fast changing disparities, geometric distortions, temporal flickering or spatial noise in the form of depth cues inconsistencies. Ideal depth  is a function of the color video for the view to be interpolated and it is used as a baseline to measure the errors in depth maps. Ideal depth  can be estimated from the rendered virtual view intensity vector $\bar{I}_{v}$, the distortion-free view intensity vector $\bar{I}_{o}$, the received depth map $\bar{Z}$ vector, focal length $F_{v}$, relative location of the rendered view $s$ ($+1$ for right and $-1$ for left), scaling factor $\alpha$ and the baseline $B$ as follows:
\vspace{-0.1in}
\begin{equation}\label{eq:Ideal}
    \bar{Z}_{IDEAL} \approx \frac{sF_{v}B}{\alpha(\bar{I}_{o} - \bar{I}_{v})+ s \frac{F_{v}B}{\bar{Z}}}
\vspace{-0.1in}
\end{equation}

 We define $\Delta \textbf{Z}$ as the difference between the \emph{ideal} depth and \emph{received} depth. Since we defined $\Delta \textbf{Z}$, distortion metrics can be formulated as follows:

\begin{itemize}
\vspace{-0.1in}
\item	\emph{Spatial Outliers (SO):} Non-zero values of $\Delta \textbf{Z}$ with non-uniform distribution results in relocation of pixels during the wrapping process. As a consequence, visual effects of these errors are spatially noticeable. Therefore, \emph{SO} is a function of $\Delta \textbf{Z}$ and can be expressed as the standard deviation of depth map errors.
    \vspace{-0.1in}
    \begin{equation}\label{eq:SO}
    \textbf{SO} = STD(\Delta \textbf{Z})
    \vspace{-0.10in}
    \end{equation}
\item	\emph{Temporal Outliers:} Temporal variation of depth map errors leads to visual distortions that can appear as impulsive intensity changes around textured region and flickering around flat regions. To take into account these temporal variations, we can express \emph{TO} as standard deviation of two depth map errors in time domain.
    \vspace{-0.1in}
    \begin{equation}\label{eq:TO}
    \textbf{TO} = STD(\Delta \textbf{Z}_{K}- \Delta \textbf{Z}_{K-1})
    \vspace{-0.10in}
    \end{equation}

\item	\emph{Temporal Inconsistencies:} Excessive and fast changing disparities result in visual distortions which can be modeled as the standard deviation of the difference of two depth values at different time instances.
    \vspace{-0.1in}
    \begin{equation}\label{eq:TI}
    \textbf{TI}=STD(\textbf{Z}_{K}-\textbf{Z}_{K-1})
    \vspace{-0.10in}
    \end{equation}
\end{itemize}

\noindent We combine these distortion measures into a single 3D vision-based quality metric for \emph{S3D} videos as follows:
\vspace{-0.1in}
\begin{equation}\label{eq:3VQM}
    \textbf{3VQM}=K(1-\textbf{SO}(\textbf{SO} \cap \textbf{TO}) )^a (1-\textbf{TI})^b (1-\textbf{TO})^c,
%\vspace{-0.10in}
\end{equation}

\noindent where $K$ is a scale factor that we choose to be 5; and the constants $a$, $b$, and $c$ are determined empirically. In~\cite{Solh2011}, we suggested to use the following values: $a=8$, $b=8$ and $c=6$.

\begin{figure*} [!t]
%\vspace{0.3cm}
%\begin{tabular}{c}

  \centering
\begin{minipage}[b]{0.25\textheight}
  \centering
\includegraphics[width=0.9\linewidth]{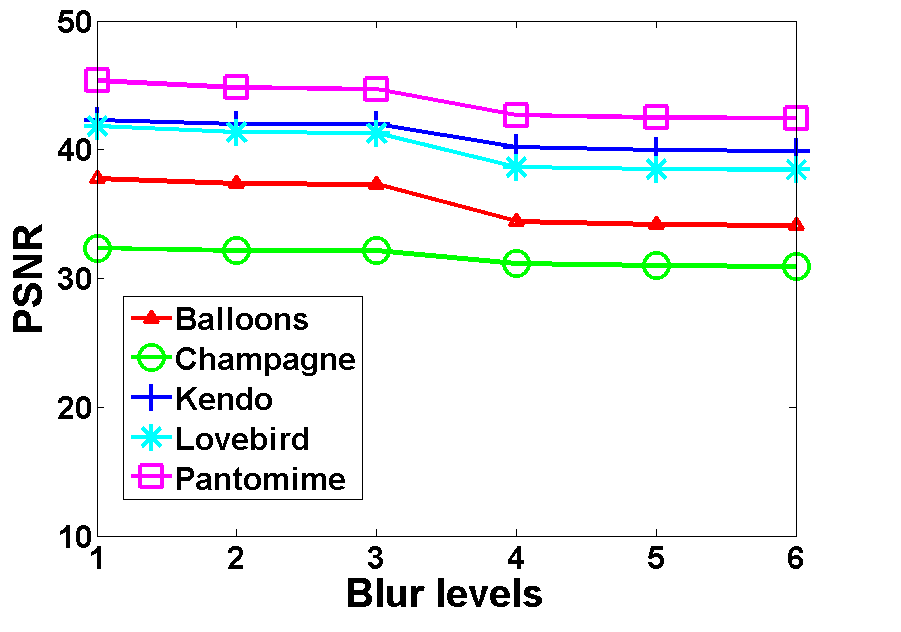}
  \centerline{\footnotesize{(a) Blurred depth videos-\emph{PSNR} }}%\medskip
\end{minipage}
  \centering
\begin{minipage}[b]{0.25\textheight}
  \centering
\includegraphics[width=0.9\linewidth]{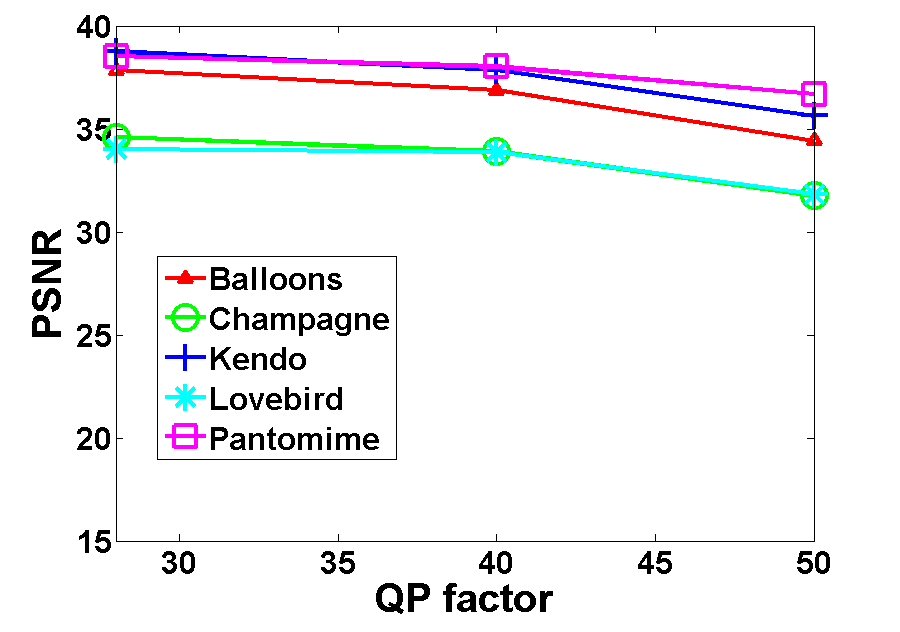}
  \centerline{\footnotesize{(e) Compressed depth videos-\emph{PSNR} }}%\medskip
\end{minipage}
  \centering
\begin{minipage}[b]{0.25\textheight}
  \centering
\includegraphics[width=0.9\linewidth]{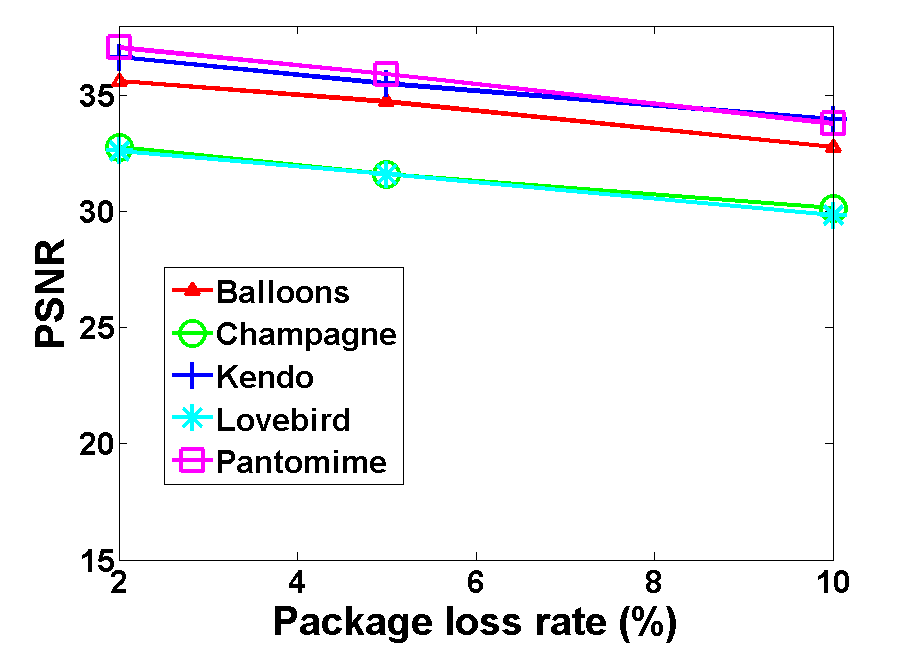}
  \centerline{\footnotesize{(i) Transmitted depth videos-\emph{PSNR} }}%\medskip
\end{minipage}
  \centering
\begin{minipage}[b]{0.25\textheight}
  \centering
\includegraphics[width=0.9\linewidth]{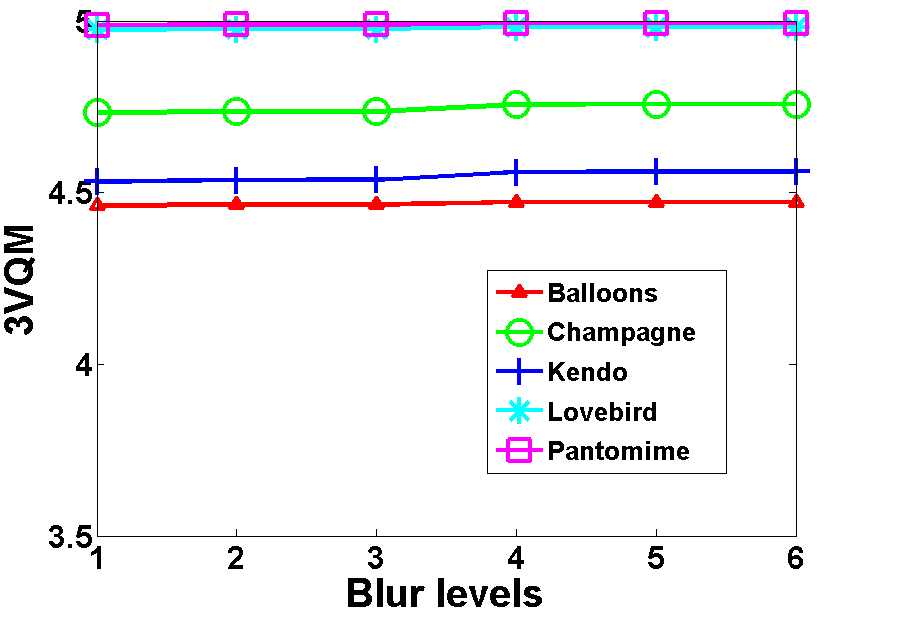}
  \centerline{\footnotesize{(b) Blurred depth videos-\emph{3VQM} }}%\medskip
\end{minipage}
  \centering
\begin{minipage}[b]{0.25\textheight}
  \centering
\includegraphics[width=0.9\linewidth]{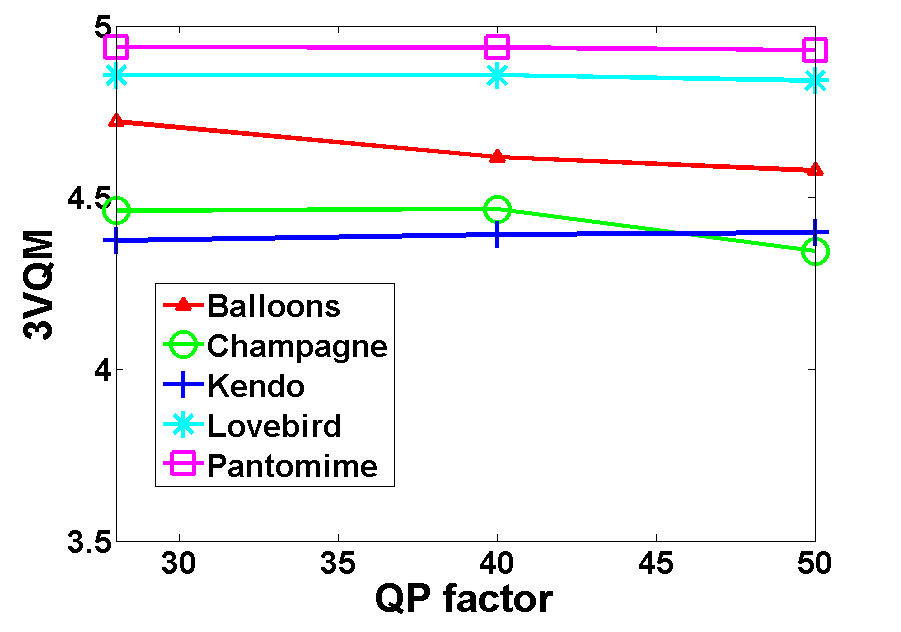}
  \centerline{\footnotesize{(f) Compressed depth videos-\emph{3VQM} }}%\medskip
\end{minipage}
  \centering
\begin{minipage}[b]{0.25\textheight}
  \centering
\includegraphics[width=0.9\linewidth]{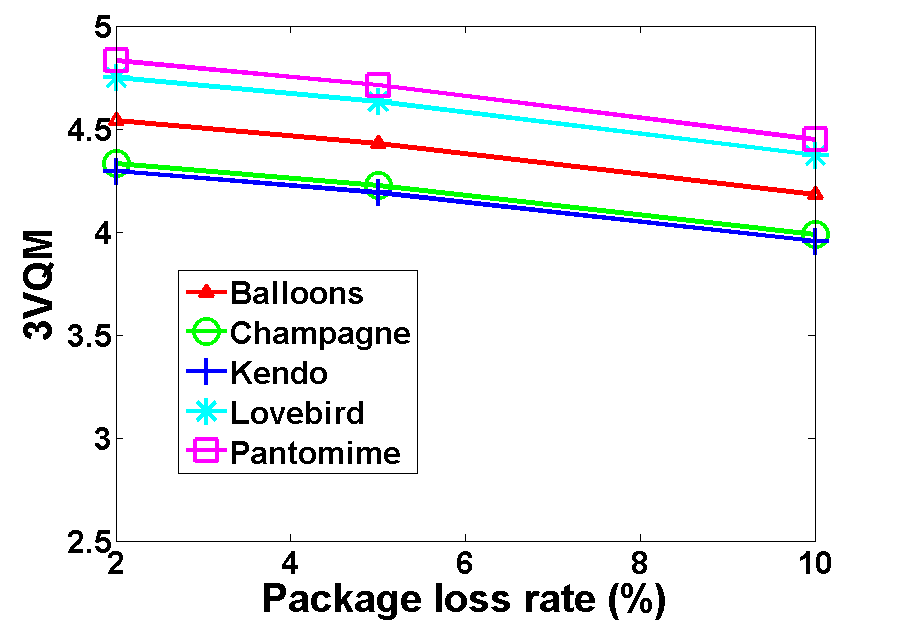}
  \centerline{\footnotesize{(j) Transmitted depth videos-\emph{3VQM} }}%\medskip
\end{minipage}
  \centering
\begin{minipage}[b]{0.25\textheight}
  \centering
\includegraphics[width=0.9\linewidth]{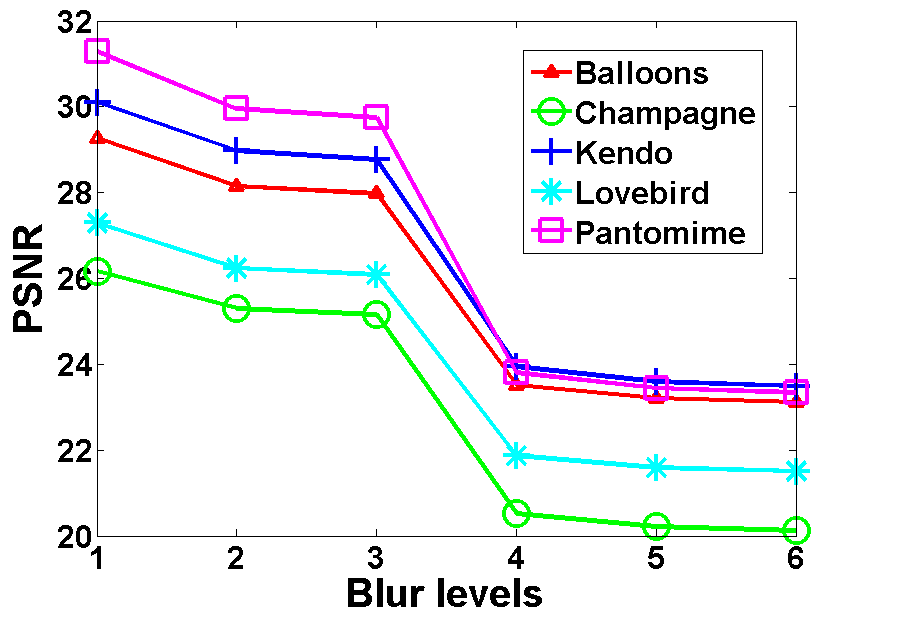}
  \centerline{\footnotesize{(c) Blurred color videos-\emph{PSNR} }}%\medskip
\end{minipage}
  \centering
\begin{minipage}[b]{0.25\textheight}
  \centering
\includegraphics[width=0.9\linewidth]{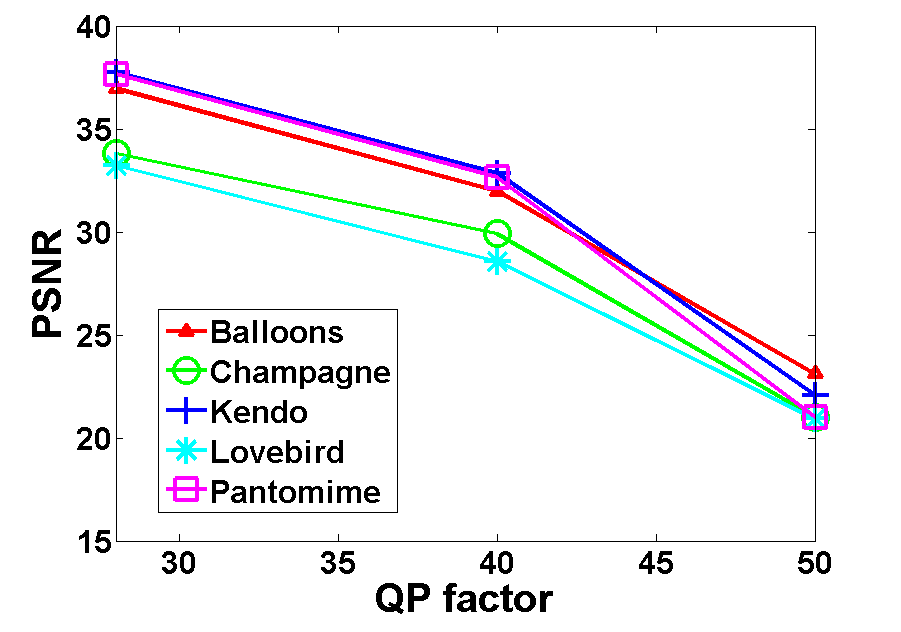}
  \centerline{\footnotesize{(g) Compressed color videos-\emph{PSNR} }}%\medskip
\end{minipage}
  \centering
\begin{minipage}[b]{0.25\textheight}
  \centering
\includegraphics[width=0.9\linewidth]{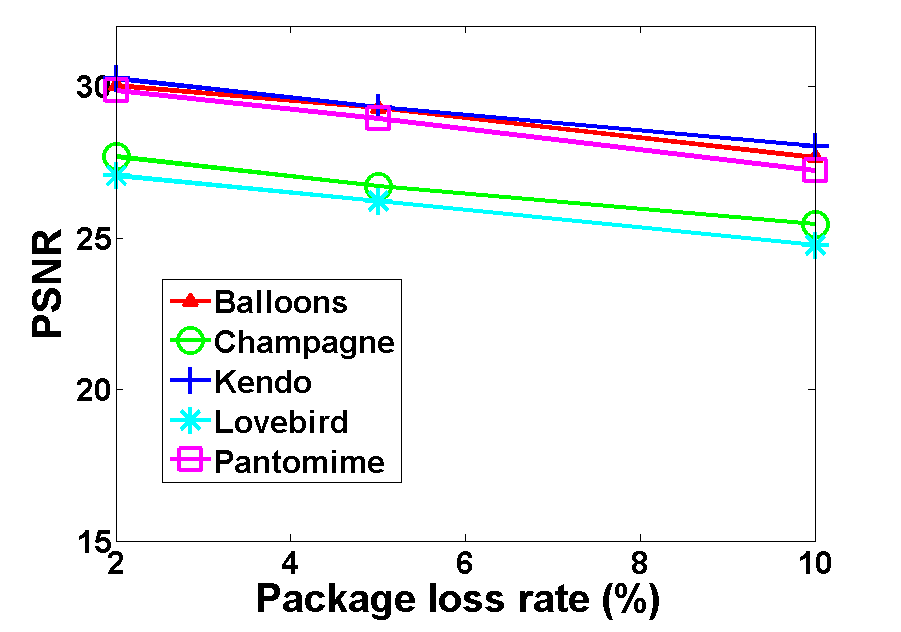}
  \centerline{\footnotesize{(k) Transmitted color videos-\emph{PSNR} }}%\medskip
\end{minipage}
  \centering
\begin{minipage}[b]{0.25\textheight}
  \centering
\includegraphics[width=0.9\linewidth]{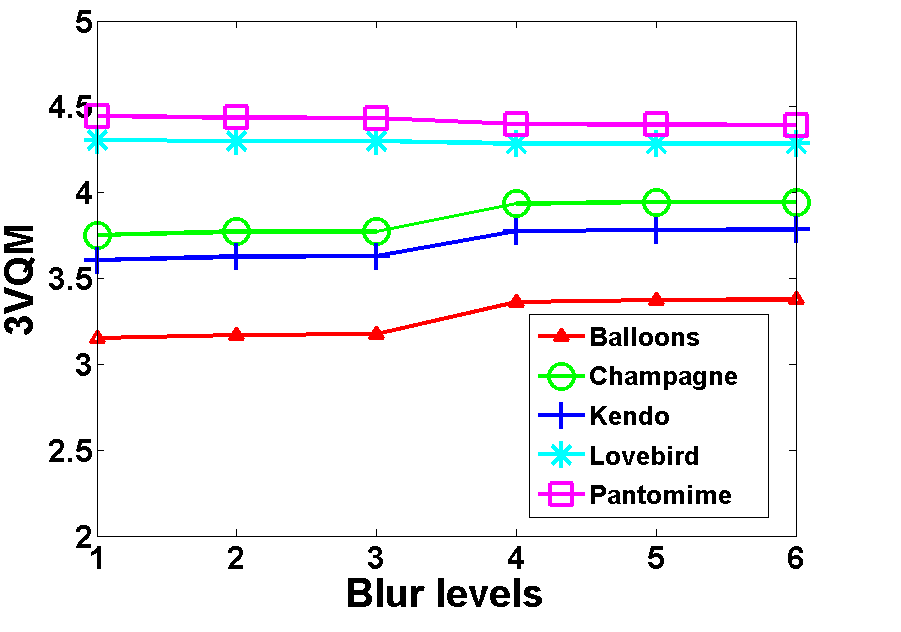}
  \centerline{\footnotesize{(d) Blurred color videos-\emph{3VQM} }}%\medskip
\end{minipage}
  \centering
\begin{minipage}[b]{0.25\textheight}
  \centering
\includegraphics[width=0.9\linewidth]{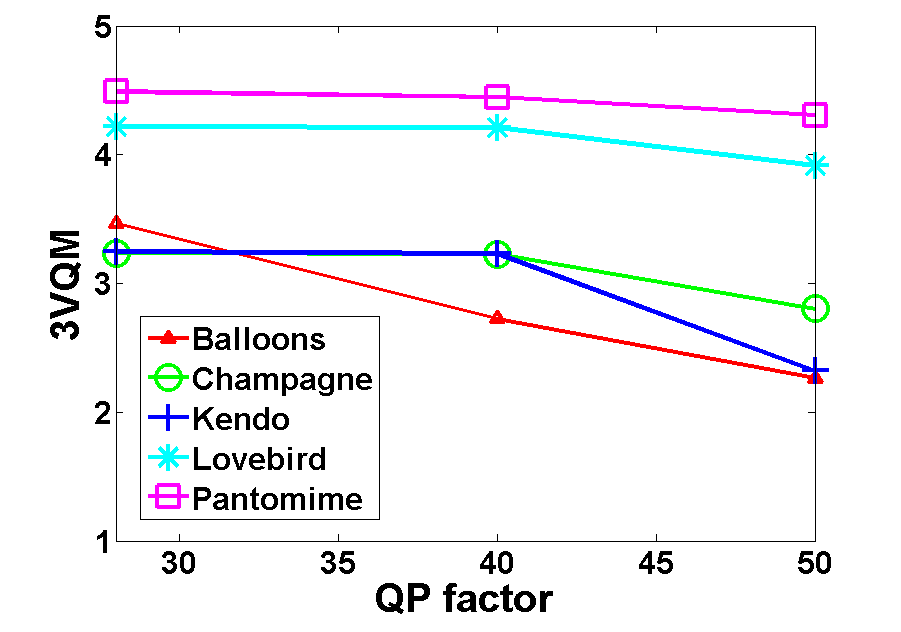}
  \centerline{\footnotesize{(h) Compressed color videos-\emph{3VQM} }}%\medskip
\end{minipage}
  \centering
\begin{minipage}[b]{0.25\textheight}
  \centering
\includegraphics[width=0.9\linewidth]{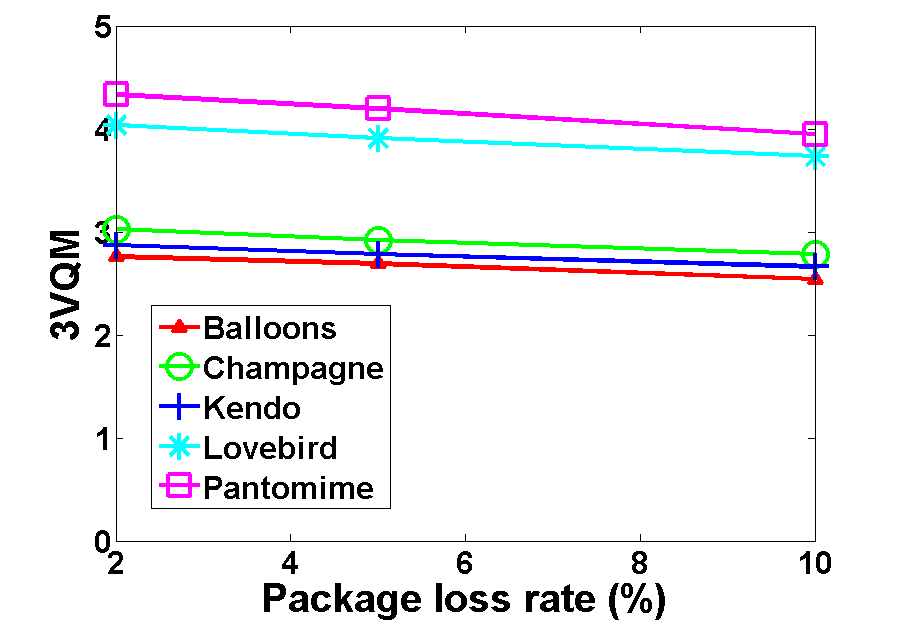}
  \centerline{\footnotesize{(l) Transmitted color videos-\emph{3VQM} }}%\medskip
\end{minipage}

\vspace{-0.1in}
\caption{\emph{3VQM} and \emph{PSNR} results}
\label{fig:RESPSNR}
\vspace{-0.2in}

\end{figure*}

\section{PERFORMANCE EVALUATION OF 3VQM}
\vspace{-0.08in}
\label{sec:performance}

In this section, we test the effectiveness of 3VQM in capturing errors and inconsistencies in the rendered depth-based 3D videos. At first, we will apply blur kernel under varying parameters to model the changes in naturalness and viewing experience as outlined in~\cite{Kaptein08}. Then, we focus on compression artifacts and transmission losses that lead to visual distortions. We use \emph{Balloons}, \emph{Champagne Tower}, \emph{Kendo}, \emph{Lovebirds} and \emph{Pantomime} sequences from \emph{3DMobile} project. Virtual views are rendered using DIBR \cite{Fehn04}. The hierarchical hole filling (HHF) is performed to avoid occlusion/disocclusion problems \cite{Solh2010}. We also render virtual views using ground truth depth maps and reference views to get a baseline for comparison. PSNR and 3VQM results for the degraded video sets are given in Fig.\ref{fig:RESPSNR} and we will refer to this figure throughout this section.

\vspace{-0.10in}
\subsection{Simulating Artifacts}
\label{ssec:blur}
We simulate the inaccuracy in depth maps using a Gaussian blur kernel. We implement different blur levels: $7\times 7$ kernel with $\sigma=2$, $\sigma=5$, $\sigma=10$  and $19\times 19$ kernel with $\sigma=10$, $\sigma=20$ and $\sigma=80$.

Compression artifacts lead to visual distortions that degrade the quality of user experience. We use \emph{ver. 18.4} of  \emph{H.264/AVC} reference software to separately encode and decode ground truth depths and color videos \cite{H2642}. We use \emph{CABAC} as entropy coding method and perform different levels of compression with $QP=28$, $QP=40$ and $QP=50$.

We packetize each frame as one packet and we use the~\emph{Gilbert Elliot} model to simulate the transmission losses. Usually, interpolation and error concealment algorithms are used to fill in the lost data and packets. In this work, we do not include any interpolation algorithm. \emph{Packet loss rate} is set to \emph{2\%}, \emph{5\%} and \emph{10\%}. We perform different realizations of color and depth videos on five video sequences and using three different packet loss ratios.

\begin{small}

\begin{table*}[t]
\centering
\caption{ Validation scores for Average PSNR, Weighted Average PSNR, Average SSIM and 3VQM.}
\begin{tabular}{c|c|c|c|c|c|c}
  \hline
   &\bf RMSE &\bf CC &\bf ROCC &\bf MAE &\bf OR &\bf \boldmath$\sigma_{DMOS}$\\
   \hline
    \textbf{Average PSNR} & 0.946 & 0.731 & 0.715 & 0.822 & 0.194 & 0.789\\   
    \textbf{Weighted Average PSNR} & 0.935 & 0.755 & 0.777 & 0.790& 0.194 & 0.789\\   
    \textbf{Average SSIM} & 0.806 & 0.598 & 0.542 & 0.621 & 0.130 & 0.789\\
   \textbf{3VQM} & 0.616 & 0.894 & 0.789 & 0.517 & 0.000 & 1.008\\
   \hline
\end{tabular}

\vspace{-0.10in}
\label{tbl:Simulations}
\end{table*}

\vspace{-0.15in}

\end{small}

\subsection{Performance Evaluation}
\label{ssec:performance}
We synthesize the virtual views using reference color videos and degraded depth maps. Also we generate synthesized views  using degraded color videos along with ground truth depth maps. We report the results where the distortion is applied to one channel, i.e., either the depth map or the reference video.

Compression and blurring of depth maps result in losing information that mostly corresponds to high frequency content or edges within the depth maps. But with depth maps having a simple structure and one color channel, compression and blurring lead to smoothing, which decreases the spatial and temporal variation. Hole filling also compensates for the inaccuracies that result from the smoothing of depth maps. The subjective quality of the rendered videos based on blurred and compressed depths are similar to the ones that are based on ground truth. PSNR decreases after a certain blur level, however 3VQM is almost insensitive to the blurring applied to the depth as it can be seen in parts \emph{a}, \emph{b}, \emph{e} and \emph{f} of Fig.\ref{fig:RESPSNR}.

 Color frames in the reference view are represented with three channels and the structure is inherently more complicated than depth. Objects located at the same distance with respect to the camera frame are represented with the same value in the depth maps. Whereas, pixel values of the same objects that are located at the same depth can significantly differ depending on the content in the color video. Thus, degradation in the color video directly degrades the rendered video. Degraded color videos result in lower PSNR and 3VQM values for most of the sequences as it is given in parts \emph{c}, \emph{g}, and \emph{h} of Fig. \ref{fig:RESPSNR}.

 3VQM values decrease as we increase the level of compression applied to color videos as in Fig.\ref{fig:RESPSNR}(h). In contrast, 3VQM gets slightly higher when  we increase the level of blur as it is shown Fig.\ref{fig:RESPSNR}(d). Since blurring smooths the image blockwise, the difference between the reference and rendered view becomes less in Eq.(\ref{eq:Ideal}) and it results in higher \emph{3VQM} because of lower $\Delta \textbf{Z}$. Although 3VQM is more sensitive to compression and blur  when it is applied to color video, the metric behaves differently with transmission losses. This is expected as with losses, the whole depth and color videos are lost and the metric value directly drops for all of the sequences as it is illustrated in Fig.\ref{fig:RESPSNR}(i-l). We packetize depth and color videos that are already compressed with H.264. \emph{QP} is set to 40 for all sequences. Transmission follows compression and errors are cumulative in the 3DTV processing chain. Thus, \emph{PSNR} and \emph{3VQM} decreases for all of the sequences after transmission.

The quantities $SO$ and $TO$ are functions of $\Delta \textbf{Z}$. Therefore, 3VQM directly depends on the reliability of ${Z}_{Ideal}$, which is expressed in Eq.(\ref{eq:Ideal}). All of the parameters except $\alpha$ are based on the DIBR configuration. However, we need to scale the difference of rendered and distortion-free views so that it will be effective in determining the value of ${Z}_{Ideal}$. If the difference term is not scaled with a reasonable value of $\alpha$ , the second term in the denominator will dominate the expression. Thus, ${Z}_{Ideal}$ will be approximately equal to the received depth map and this results in  \emph{SO}=1 and \emph{TO}=1. To scale the terms in the denominator into the same level, $\alpha$ is set to $120$ for all of the sequences.

%We expect that the temporal outlier in \emph{3VQM} would be high in this case because of the inconsistencies resulting from the missing reference video frames or the missing depth map frames.
%However, in this work we assumed that the quality of the  lost package equals to the minimum value of the objective quality metric. Since transmission follows compression, errors are cumulative and the resulting objective quality values are lower than compression as is it illustrated in Table \ref{tab:Res}.
%Probability of transferring from good state to bad state is set to \emph{0.02777} and we assigned \emph{0.25} to the probability of transferring from the bad state to the good state. Simulations are performed for \emph{50} frames and the average \emph{PSNR} and \emph{3VQM} values are calculated for the overall quality.

%
%\vspace{-0.10in}
\section{Validation of 3VQM}
\label{ssec:validation}
\vspace{-0.05in}
In addition to the videos from Mobile3D project,  we captured stereo videos using Point Grey's BumbleBee2 camera. To simulate the degradation of quality, we perform H.264 based compression and estimated depth maps using stereo matching and 2D to 3D conversion methods instead of using ground truth depth map. We performed subjective quality assessment  according to the requirements mentioned in \cite{Chen2010}. Subjects evaluated the quality of video sequences with a discrete rating. Raw scores were collected and processed to give Difference Mean Opinion Scores (\emph{DMOS}).
 21 video sequences of 30 seconds length with both reference and distorted videos were used in the subjective test. Performance of \emph{3VQM} is compared with PSNR, weighted average PSNR \cite{Ozbek07:St} and structural similarity index \emph{(SSIM)} \cite{Wang04:SSIM}. A more comprehensive analysis of \emph{3VQM} including full-reference and no-reference approaches was submitted as a journal publication.

Validation scores are selected according to the VQEG recommendations. We use Root Mean Squared Error (RMSE), Pearson Linear Correlation Coefficient (CC), Spearman Rank Order Correlation Coefficient (ROCC), Mean Absolute Error (MAE), Outlier Ratio (OR) and the standard deviation of the DMOS values ($\sigma_{DMOS}$).We define outliers as the points whose distance from the reference is greater than twice the \emph{DMOS} standard deviation. High \emph{CC} and \emph{ROCC} shows coherency, low \emph{RMSE}  and \emph{MAE} indicates accuracy and  low \emph{OR} represents consistency. As it is given in Table \ref{tbl:Simulations}, 3VQM is the most accurate, coherent, and consistent among all objective measures represented in this paper.

%\vspace{-0.10in}
\section{Conclusion}
\label{sec:conc}

We evaluated the effectiveness of \emph{3VQM} in capturing depth inconsistencies by simulating compression artifacts, transmission losses and depth map estimation errors. According to the simulation results, \emph{3VQM} captures the depth inconsistencies based on errors in the reference views more effectively than errors in the depth map. Errors based on smoothing are not considered as degradation since they lead to decrease in temporal and spatial variations. We performed subjective quality assessment to validate 3VQM and we showed that 3VQM is the most accurate, coherent, and consistent among all objective measures represented in this paper.

% -------------------------------------------------------------------------
%\begin{small}
%\bibliographystyle{IEEEbib}
%\bibliography{strings,refs}
%\end{small}

\end{document}